\documentclass[sigconf]{acmart}

\usepackage{amsmath,amssymb,amsfonts}
\usepackage{algorithmic}
\usepackage{multirow}
\usepackage{bm}
\usepackage{algorithm}
\usepackage{mathtools}
\usepackage{subfigure}

\DeclarePairedDelimiter{\ceil}{\lceil}{\rceil}
\DeclareMathOperator*{\argmin}{arg\,min}

\AtBeginDocument{%
  \providecommand\BibTeX{{%
    \normalfont B\kern-0.5em{\scshape i\kern-0.25em b}\kern-0.8em\TeX}}}

\copyrightyear{2020}
\acmYear{2020}
\setcopyright{acmlicensed}\acmConference[CIKM '20]{Proceedings of the 29th ACM International Conference on Information and Knowledge Management}{October 19--23, 2020}{Virtual Event, Ireland}
\acmBooktitle{Proceedings of the 29th ACM International Conference on Information and Knowledge Management (CIKM '20), October 19--23, 2020, Virtual Event, Ireland}
\acmPrice{15.00}
\acmDOI{10.1145/3340531.3411942}
\acmISBN{978-1-4503-6859-9/20/10}
\begin{document}
\fancyhead{}

\title{RelSen: An Optimization-based Framework for Simultaneously Sensor Reliability Monitoring and Data Cleaning}

\author{Cheng Feng, Xiao Liang, Daniel Schneegass, PengWei Tian}
\affiliation{%
  \institution{Siemens Corporate Technology}
  \streetaddress{No.7, Wangjing Zhonghuan Nanlu}
  \city{Beijing}
  \country{China}}
\email{{cheng.feng, liang.xiao, daniel.schneegass, pengwei.tian}@siemens.com}

\renewcommand{\shortauthors}{Feng and Liang, et al.}

\begin{abstract}
Recent advances in the Internet of Things (IoT) technology have led to a surge on the popularity of sensing applications. As a result, people increasingly rely on information obtained from sensors to make decisions in their daily life. Unfortunately, in most sensing applications, sensors are known to be error-prone and their measurements can become misleading at any unexpected time. Therefore, in order to enhance the reliability of sensing applications, apart from the physical phenomena/processes of interest, we believe it is also highly important to monitor the reliability of sensors and clean the sensor data before analysis on them being conducted. Existing studies often regard sensor reliability monitoring and sensor data cleaning as separate problems. In this work, we propose RelSen, a novel optimization-based framework to address the two problems simultaneously via utilizing the mutual dependence between them. Furthermore, RelSen is not application-specific as its implementation assumes a minimal prior knowledge of the process dynamics under monitoring. This significantly improves its generality and applicability in practice. In our experiments, we apply RelSen on an outdoor air pollution monitoring system and a condition monitoring system for a cement rotary kiln. Experimental results show that our framework can timely identify unreliable sensors and remove sensor measurement errors caused by three types of most commonly observed sensor faults.
\end{abstract}





\maketitle

\section{Introduction}
With the trend of IoT, sensors are becoming ubiquitous. The measurements from sensors have become an important source of knowledge to decision making for both human-beings and computing machines in different domains, such as industrial process control \cite{lu2015real}, air pollution monitoring \cite{kumar2015rise} and traffic flow measurement \cite{li2006application}. Nevertheless, it is also well known that measurements from sensors (especially commodity sensors which are widely used in IoT applications) can be erroneous caused by hardware or software faults \cite{sharma2010sensor} due to various reasons such as manufacturing imperfections, device aging, extreme ambient environment conditions and even cyber attacks \cite{zhang2017cyber}. This has motivated many researchers in both industrial and academic communities to craft specialized data cleaning techniques to mitigate the effects of sensor measurement errors. These techniques can be broadly categorized into three classes:  1) state estimation methods which estimate the states of monitored processes by utilizing the prior knowledge about the process state transition dynamics and the distribution of measurement noise. The family of Bayesian filters  \cite{jazwinski1970stochastic,julier2004unscented,chui2017kalman} are within this class; 2) parameter estimation methods which estimate the parameters of the sensor measurement errors that best describe the observed sensor measurements via the learned sensor models, examples see \cite{yuan2007bayesian,wang2012truth,wen2013assessing}; 3) multi-sensor fusion techniques which combine measurements from redundant sensors to achieve improved measuring accuracy than that could be achieved by the use of a single sensor alone. Typical fusion methods combine measurements from multiple sensors using mean, median or weighted average statistics based on known noise variance or covariance of sensors \cite{sun2017multi,xiao2005scheme}. Despite many successful applications of the above three techniques in the past decades, they all have certain limitations which restrict their applicability in the IoT context. For state estimation methods, a system identification step \cite{ljung1999system} is often required to build up the mathematical models of the process dynamics before they can be applied. This step is however generally very challenging in practice. For parameter estimation methods, a training phase is often required to capture the "profile"  of the sensor dynamics via models trained by a certain amount of observed data. As a result, their performances are likely to deteriorate when the underlying process characteristics change which means that the trained models cannot well represent the sensor dynamics any more. For multi-sensor fusion techniques, on one hand sensor redundancy is not always affordable, on the other hand except the naive mean and median method, how to assign weights to redundant sensors for fusion often becomes a state estimation or parameter estimation problem \cite{li2015survey}.

Another problem we investigate is sensor reliability monitoring. We believe that monitoring the reliability of sensors can bring tremendous benefits. To name a few, sensor reliability provides an important metric for benchmarking between different sensor vendors, and customers knowing how reliable a sensor is can better decide whether to use or buy it. Furthermore, by monitoring the reliability of sensors, predictive maintenance of sensor systems can be conducted by timely identifying and replacing unreliable sensors and wrong decisions caused by misleading measurements can be avoided. Most importantly, knowing the reliability of sensors can also improve the accuracy of data cleaning by giving less weights to unreliable sensors for estimating the ground truth of the measured signals. 

In this paper we propose RelSen, an optimization-based framework for simultaneous sensor reliability monitoring and data cleaning. In RelSen every sensor is assigned a reliability score that can be updated dynamically based on the sensor's latest measurement errors in a sliding window. The reliability scores are then utilized to remove the sensor measurement errors. Specifically, we formulate both the reliability scores of sensors and the ground truth of measured signals as variables to learn by solving optimization problems only given the observed measurements. Notably, in RelSen we do not assume the dynamics of the underlying monitored processes to be predefined, which means that the system identification step as in the filtering-based state estimation methods as well as the model training phase in the parameter estimation methods are not required. This significantly improves the generality and applicability of RelSen in practice, especially in the IoT context where many processes with highly unpredictable dynamics need to be measured. To demonstrate its effectiveness, we apply RelSen for sensor reliability monitoring and data cleaning in two sensor systems: one deployed for outdoor air pollution monitoring, the other for condition monitoring of cyclones and decomposition furnaces in a cement rotary kiln. With no prior knowledge about the process dynamics of the two systems, experimental results show that our framework can timely identify unreliable sensors and outperform several existing data cleaning methods under three types of commonly observed sensor faults.

The remaining part of this paper is organized as follows. We first discuss the related work in the next section. In Section~\ref{sec:ps}, we outline the research problem of this paper. This is followed by the presentation of the RelSen framework in Section~\ref{sec:framework}. Technical implementation issues of RelSen are discussed in Section~\ref{sec:implement}. Then, the experiments on the outdoor air quality monitoring system and the cement rotary kiln condition monitoring system are presented in Section~\ref{sec:airquality} and~\ref{sec:kiln} respectively. Finally, we draw the conclusion and discuss possible extensions in the last section.

\section{Related Work}
Sensor reliability evaluation and data cleaning are generally regarded as two separate but closely related tasks. Sensor reliability or accuracy is commonly evaluated by some summary statistics of the sensor measurement errors. However, as the ground truth of the measured signals is generally unknown, sensor data cleaning methods have to be applied beforehand. Among the common approaches for sensor data cleaning are state estimation methods \cite{jazwinski1970stochastic,del1996non,van2001unscented,julier2004unscented}, parameter estimation methods such as Bayesian estimation algorithms \cite{yuan2007bayesian,yuan2008robust} and truth estimation algorithms \cite{wang2012truth}, and multi-sensor fusion techniques \cite{wen2004fuzzy,gustafsson2010statistical}. To formulate the whole workflow, the authors of \cite{wen2014accuracy} proposed a sensor accuracy estimation framework which consists of four layers: pre-processing, state estimation, accuracy estimation and accuracy indexing. In their work, several taxonomies are proposed for the methods that can be used to implement data cleaning.

In RelSen, the two problems (sensor reliability monitoring and data cleaning) are jointly tackled in a single framework and their mutual dependence is explicitly utilized. Similarly, the authors of \cite{zhang2014cleaning} also proposed a sensor reliability-based data cleaning method for environmental sensing applications. In their method named Influence Mean Cleaning (IMC), the reliability score of a sensor is incrementally updated by checking the distance between its measurement and the predicted true state of the underlying monitored process. The reliability score of a sensor with a distance smaller than a user-defined threshold will be increased, otherwise the reliability score will be decreased. The true state of the underlying monitored process is calculated as the sensor reliability-weighted mean of measurements from a group of spatially correlated sensors. Furthermore, the authors also discussed the effect of removing unreliable sensors on the accuracy of data cleaning in \cite{zhang2016reduce}. Compared with IMC, RelSen conducts sensor reliability-based data cleaning with an optimization-based framework which has better  interpretability both intuitively and mathematically.

It is common that the same phenomenon can have many different views from different entities. To discover the truth from multiple data sources, entity reliability-based truth discovery has been studied for many years in the information retrieval domain \cite{li2016survey}. Among the commonly used methods for truth discovery in information retrieval are voting-based methods \cite{pasternack2010knowing,pasternack2011making}, optimization-based methods \cite{li2014confidence,li2014resolving} and probabilistic inference-based methods \cite{zhao2012probabilistic,pasternack2013latent,li2015discovery}. RelSen also takes an optimization-based method for sensor data cleaning, however, the problem we face is more complex as we need to consider the evolving dynamics of the monitored processes, the unexpected faults from sensors and the processes whose states are only reported by a single data source (sensor).

\section{Problem Statement}
\label{sec:ps}
In this section, we formulate the problem class which we study in this paper. We consider the general case where there are a number of sensors monitoring multiple physical processes in a system. Let $\mathcal{P}$ be the set of physical processes, $\mathcal{S}$ be the set of sensors in the system, we use $\mathcal{S}_p$ to denote the set of sensors that are monitoring the physical process $p$, where $p \in \mathcal{P}$, $|\mathcal{S}_p| \geq 1$ and $\sum_{p \in \mathcal{P}} |\mathcal{S}_p| = |\mathcal{S}|$. That is to say, each physical process is monitored by one or more sensors, and each sensor can only monitor one physical process. Furthermore, 
we assume that the monitored processes are cross-correlated which is common in practice, and there is no prior knowledge about the state transition dynamics of the monitored processes.

Let $\mathbf{x}^t = [x^t_1,x^t_2,\ldots,x^t_{|\mathcal{S}|}]$ be the measurements from the sensors at a given discrete time $t$, where the timestamps $t = 1:N$ are a totally ordered set. Assume a physical process $p$ is measured by the sensor $s$, we define $x^t_s = z^t_p + e^t_s$ in which $z^t_p$ is the hidden state of process $p$, $e^t_s$ is the sensor measurement error. Our target is to quantify and monitor $\mathbf{c}^t = [c^t_1,c^t_2,\ldots,c^t_{|\mathcal{S}|}]$ which are the reliability scores of sensors and remove sensor measurement errors to reveal the ground truth of monitored process states $\mathbf{z}^t = [z^t_1,z^t_2,\ldots,z^t_{|\mathcal{P}|}]$.

\section{RelSen Framework}
\label{sec:framework}
\begin{figure}
  \centering
  \includegraphics[width=.99\linewidth]{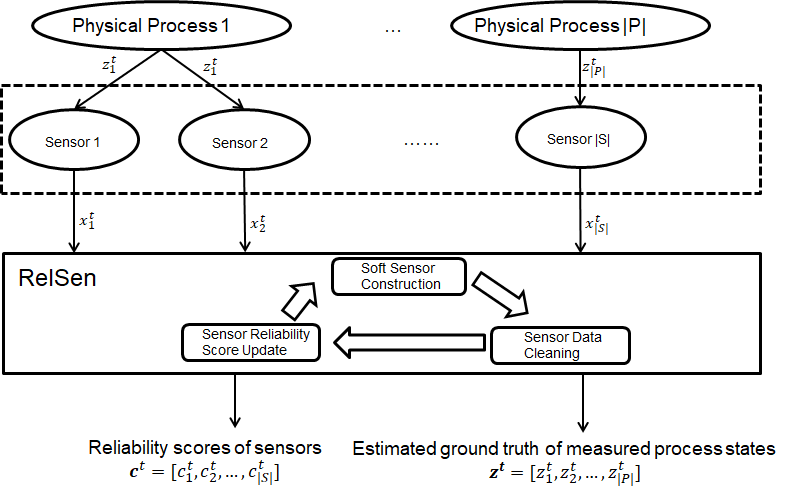}
\caption{The schematic diagram of the RelSen framework}
  \label{fig:framework}
\end{figure}

In this section we present the RelSen framework to address the problems described in the previous section. The schematic diagram of RelSen is illustrated in Figure~\ref{fig:framework}. Specifically, the framework consists of three modules: automatic soft sensor construction, sensor data cleaning and sensor reliability score update. With sensor measurements as the only input, these three modules run iteratively to calculate the reliability scores of sensors and estimate the ground truth of measured process states in real-time. In the remainder of this section, the implementation of each module will be described in detail.

\subsection{Automatic Soft Sensor Construction}
\label{sec:localreg}
The goal of this module is to build soft sensors to provide extra information for sensor data cleaning and sensor reliability score update by utilizing the correlation between multiple processes. 

The soft sensors are automatically constructed by fitting \emph{random local} linear regression models. Concretely, let $p$ be a target physical process. To build up a soft sensor for $p$ at time $t$, we first randomly select $\ceil{r\times |\mathcal{S} \setminus \mathcal{S}_p|}$ sensors from the sensor set $\mathcal{S}\setminus \mathcal{S}_p$, where $r\in (0,1]$ is a tunable ratio defined by the user. Furthermore, let $\mathcal{S}^t_{p,m}$ be the selected sensors which we call \emph{explanatory sensors} for setting up a soft sensor $m$ for process $p$ at time $t$, $\mathbf{x}^{t}_{p,m}$ be the vector consists of the measurements from the explanatory sensors, we define a neighbor set $\mathcal{J}^t_{p,m}$ for the point $\mathbf{x}^{t}_{p,m}$. The neighbor set is derived as the K-nearest neighbors using Euclidean distance from the set of observed measurements in the time interval $[1:t-1]$. Then the signal of the soft sensor $m$ for process $p$ at time $t$ is given by:
\begin{equation}
\label{eq:softsensor}
y^t_{p,m} = \mathbf{w}^t_{p,m} \mathbf{x}^{t}_{p,m} + b^t_{p,m}  = \sum_{s \in \mathcal{S}^t_{p,m}} w^t_{p,m,s} x^t_{s} + b^t_{p,m} 
\end{equation}
%Let  $\mathbf{w}^t_{p} = [] \mathbf{x}^t_{\mathcal{S}^t_{p,m}} + w^t_{p,m,0}$
where $w^t_{p,m,s}$, the weight of an explanatory sensor $s$ for constructing the soft sensor $m$ for process $p$ at time $t$, takes the solution of the following optimization problem:
\begin{equation}
\label{eq:fitsoft}
\argmin_{\mathbf{w}^t_{p,m},b^t_{p,m}} \quad \sum_{\mathbf{x}^{t'}_{\mathcal{S}^t_{p,m}} \in \mathcal{J}^t_{p,m} } (z^{t'}_p - \mathbf{w}^t_{p,m} \mathbf{x}^{t'}_{\mathcal{S}^t_{p,m}} - b^t_{p,m})^2
\end{equation} in which $\mathbf{x}^{t'}_{\mathcal{S}^t_{p,m}}$ denotes the vector consisting of the measurements from the sensor set $\mathcal{S}^t_{p,m}$ at time $t'$. Moreover, we denote $E_{p,m}^t$ as the fitting error of the soft sensor, such that
\begin{equation}
E_{p,m}^t=\frac{1}{K}\sum_{\mathbf{x}^{t'}_{\mathcal{S}^t_{p,m}} \in \mathcal{J}^t_{p,m} } (z^{t'}_p - \mathbf{w}^t_{p,m} \mathbf{x}^{t'}_{\mathcal{S}^t_{p,m}} - b^t_{p,m})^2 \nonumber
\end{equation}where $K$ is the size of the neighbor set defined by the user. The weights of explanatory sensors and the fitting error will be used to define the reliability score of the soft sensor which will be shown in the next subsection.

Furthermore, at each time step, we construct $M_p$ soft sensors for physical process $p$ using the above method, where $M_p \geq 0$ is defined by the user. The soft sensors constructed by fitting \emph{random local} linear regression models instead of traditional linear regression models enjoy two important properties: 1) the soft sensors are weakly correlated with each other as they are fitted using different set of explanatory sensors and different data points, thus the prediction errors they make tend to be uncorrelated; 2) because of the usage of local regression, the soft sensors are able to capture nonlinear correlation between multiple processes and can be promptly adapted when the process characteristics change \cite{zhu2011local,kano2012virtual}. 

\subsection{Sensor Data Cleaning}
The target of this module is to remove sensor measurement errors and estimate the ground truth of process states by utilizing the reliability scores of sensors. Specifically, assuming the sensor reliability scores $\mathbf{c}^{t}$ are known positive constants in this module, we propose to reveal the ground truth of process states by solving the following optimization problem:
\begin{eqnarray*}
\underset{\mathbf{z}^t}{\min} \quad L_1 &=& \sum_{p \in \mathcal{P}} \sum_{s \in \mathcal{S}_p} c_s^{t} (z_p^t-x_s^t)^2 \\
&& +\sum_{p \in \mathcal{P}} \sum_{m=1}^{M_p} c^{t}_{p,m} (z_p^t-y^t_{p,m})^2 \\
&& + \sum_{p \in \mathcal{P}} \gamma_p (z_p^t-z_p^{t-1})^2 \\
\end{eqnarray*}in which $c^{t}_{p,m}$ is the reliability score of the soft sensor $m$ for process $p$ computed as follows:
\begin{eqnarray*}
c^{t}_{p,m} = \frac{\sum_{s \in \mathcal{S}^t_{p,m}} |w^t_{p,m,s}| c_s^{t} }{\sum_{s \in \mathcal{S}^t_{p,m}} |w^t_{p,m,s}|} (1-e^t_{p,m})
\end{eqnarray*}where $e^t_{p,m} \in [0,1]$ is the normalized fitting error of the soft sensor, such that $e^t_{p,m}=\frac{E_{p,m}^t-\min(\mathbf{E})}{\max(\mathbf{E})-\min(\mathbf{E})}$ in which $\mathbf{E}$ is the set of fitting errors for all constructed soft sensors until $t$, such that 
\begin{equation*}
    \mathbf{E}=\{E_{p',m'}^{t'}\} \quad \forall p' \in \mathcal{P}, m' \in [1:M_p], t' \in [1:t].
\end{equation*}Intuitively, the reliability score of a soft sensor is defined as the weighted sum of the reliability scores of its explanatory sensors scaled by the normalized fitting error when constructing the soft sensor. In this way, an explanatory sensor with a larger absolute weight in constructing the soft sensor contributes a larger proportion of its reliability score to the soft sensor, and a soft sensor with a higher fitting error will have a lower reliability score.

The motivation behind the loss function $L_1$ is as follows: 1) The first term measures sensor reliability weighted distance
between measurements from the sensors with the ground truth of monitored process states. 2) The second term measures sensor reliability weighted distance between outputs from the constructed soft sensors with the ground truth of monitored process states.
By minimizing the above two terms, the estimated ground truth of process states will be closer to the signals from more reliable sensors. 3) The third term is a smoothing factor where $\gamma_p$ is a user-defined hyperparameter which controls the smoothness for process $p$.

Since $L_1$ is convex, by making the derivative with respect to $z_p^t$ be zero, we get a closed form solution:
\begin{eqnarray}
\label{eq:statestimation}
z_p^t= \frac{ \sum_{s \in \mathcal{S}_p} c_s^{t} x_s^t + 
\sum_{m=1}^{M_p} c^t_{p,m} y^t_{p,m} + \gamma_p z_p^{t-1} }{  \sum_{s \in \mathcal{S}_p} c_s^{t} 
+\sum_{m=1}^{M_p} c^t_{p,m} + \gamma_p  } \quad \forall p \in \mathcal{P}
\end{eqnarray}Intuitively, the solution indicates that more reliable sensors have larger weights in estimating the ground truth of process states.

\subsection{Sensor Reliability Score Update}
\label{sec:score_update}
The goal of this module is to update reliability scores of sensors based on their latest measurement errors assuming that the ground truth of monitored process states are known constants. Specifically, let $l$ be the length of a sliding window, we propose to update sensor reliability scores by solving the following constrained optimization problem:
\begin{eqnarray*}
\underset{\mathbf{c}^t}{\min} \quad L_2 &=& \sum_{p \in \mathcal{P}} \sum_{s \in \mathcal{S}_p} \sum_{k=t-l}^t c_s^t (z_p^k-x_s^k)^2 \\
&& + \sum_{p \in \mathcal{P}} \sum_{m=1}^{M_p} \sum_{k=t-l}^{t} c_{p,m}^k (z_p^k-y^k_{p,m})^2 \\
 &\text{s.t.}& \quad \sum_{s \in \mathcal{S}} \exp({-c_s^t}) = 1 
\end{eqnarray*}where
\begin{eqnarray*}
c_{p,m}^k = \frac{\sum_{s \in \mathcal{S}^k_{p,m}} |w^k_{p,m,s}| c_s^t }{\sum_{s \in \mathcal{S}^k_{p,m}} |w^k_{p,m,s}|} (1-e^k_{p,m})
\end{eqnarray*}Specifically, by minimizing the reliability score weighted distance between the sensor signals and the estimated ground truth of monitored process states as in the loss function $L_2$, we assign larger reliability scores to sensors with smaller measurement errors in the sliding window. A constraint term is also required to make the optimization problem bounded. We choose to constrain the sum of exponential of negative reliability scores to be 1. This particular form has a nice property that reliability scores are also guaranteed to be positive without additional constraint terms needed. Theoretically, other forms of constraints are also allowed.

To solve the constrained optimization problem, we introduce a Lagrange multiplier $\lambda$ for the constraint. Then we obtain the following Lagrangian:
\begin{eqnarray*}
L(\mathbf{c}^t, \lambda) &=&  \sum_{p \in \mathcal{P}} \sum_{s \in \mathcal{S}_p} \sum_{k=t-l}^t c_s^t (z_p^k-x_s^k)^2 \\
&& + \sum_{p \in \mathcal{P}} \sum_{m=1}^{M_p} \sum_{k=t-l}^{t} c_{p,m}^k (z_p^k-y^k_{p,m})^2 \\
&&+  \lambda \Big( \sum_{s \in \mathcal{S}} \exp({-c^t_s})-1 \Big) 
\end{eqnarray*}
Since that the above function is convex, the global optimum can be achieved by making partial derivative with respect to $c^t_s$ be zero and we obtain:
\begin{eqnarray}
\label{eq:partial}
\sum_{k=t-l}^t (z_p^k-x_s^k)^2 + \sum_{p' \in \mathcal{P} \setminus p} \sum_{m=1}^{M_{p'}} \sum_{k=t-l}^t g_{p',m,s}^k (z_{p'}^k-y^k_{p',m})^2 \nonumber  \\ 
= \lambda \exp(-c_s^t)
\end{eqnarray}where
\begin{eqnarray*}
g_{p',m,s}^k = \frac{ \mathbf{1}(s \in \mathcal{S}^k_{p',m} ) |w^k_{p',m,s}| }{\sum_{s' \in \mathcal{S}^k_{p',m}} |w^k_{p',m,s'}|}  (1-e^k_{p',m})
\end{eqnarray*}in which $\mathbf{1}(\cdot)$ is an indicator function which equals 1 when the condition is satisfied and 0 otherwise. Moreover, since $\sum_{s \in \mathcal{S}} \exp({-c_s^t}) = 1$, we obtain:
\begin{small}
\begin{eqnarray*}
\lambda = \sum_{p \in \mathcal{P}} \sum_{\hat{s} \in \mathcal{S}_p} \sum_{k=t-l}^t \Big\{ (z_p^k-x_{\hat{s}}^k)^2 +  \sum_{p' \in \mathcal{P} \setminus p} \sum_{m=1}^{M_{p'}}  g_{p',m,\hat{s}}^k (z_{p'}^k-y^k_{p',m})^2 \Big\}
\end{eqnarray*}
\end{small}
Replacing $\lambda$ back to Equation~\ref{eq:partial}, we finally obtain:
\begin{eqnarray}
\label{eq:reliability1}
c_s^t =- \ln (f) \quad \forall s \in \mathcal{S}
\end{eqnarray}
where
\begin{small}
\begin{equation}
 f=\frac{\sum_{k=t-l}^t \Big\{ (z_p^k-x_s^k)^2 +\sum_{p' \in \mathcal{P} \setminus p} \sum_{m=1}^{M_{p'}} g^k_{p',m,s} (z_{p'}^k-y^k_{p',m})^2 \Big\} }{ \lambda} \nonumber
\end{equation}
\end{small}

To sum up, we give the algorithm for real-time sensor reliability monitoring and data cleaning in Algorithm~\ref{alg:realtime}.
\begin{algorithm}
  \caption{Real-Time Sensor Reliability Monitoring and Data Cleaning}
  \begin{algorithmic}[1]
  	\REQUIRE $l$, $r$, $K$, $\{M_p\}_{p \in \mathcal{P}}$, $\{\gamma_p\}_{p \in \mathcal{P}}$
  	\LOOP 
  	\STATE{Receive new measurements $\mathbf{x}^t$ from sensors} 
  	\FOR{$p \in \mathcal{P}$} 
  	\FOR{$m=1,\ldots,M_p$}
  	\STATE{Randomly select $\ceil{r\times|\mathcal{S} \setminus \mathcal{S}_p|}$ explanatory sensors from the sensor set $\mathcal{S}\setminus \mathcal{S}_p$}
  	\STATE{Derive the $K$-nearest neighbors $\mathcal{J}^t_{p,m}$ for the point $\mathbf{x}^{t}_{p,m}$, where $\mathbf{x}^{t}_{p,m}$ consists of measurements from the explanatory sensors}
  	\STATE{Construct soft sensor $y_{p,m}^t$ as in Equation~\ref{eq:softsensor} by solving the optimization problem in Equation~\ref{eq:fitsoft}}
  	\ENDFOR
  	\ENDFOR
  	
  	\STATE{Estimate ground truth of process states $\mathbf{z}^t$ by Equation~\ref{eq:statestimation}}
  	\STATE{Update sensor reliability scores $\mathbf{c}^t$ by Equation~\ref{eq:reliability1}}
  	\STATE{Emit $\mathbf{c}^t$ and $\mathbf{z}^t$ }
  	\ENDLOOP
  \end{algorithmic}
   \label{alg:realtime}
\end{algorithm}

\subsection{Warm-up period}
In order to run Algorithm~\ref{alg:realtime}, the sensor reliability scores and the ground truth of monitored process states in the first $T$ time steps must be derived beforehand, where $T>\max(l,K)$. Thus, we define the time interval $[1:T]$ as the warm-up period. Notably, in the warm-up period we assume the reliability score for each sensor is unchanged. Thus we use $\mathbf{c} = [c_1,c_2,\ldots,c_{|\mathcal{S}|}]$ to denote the reliability scores of sensors within this period. 

The sensor reliability scores and the estimated ground truth of monitored process states in the warm-up period are derived by solving a joint optimization problem as follows:
\begin{eqnarray*}
\underset{\mathbf{c}, \{\mathbf{z}^t\}_{t=1}^T}{\min}  L &=& \sum_{p \in \mathcal{P}} \sum_{s \in \mathcal{S}_p} \sum_{t=1}^T c_s (z_p^t-x_s^t)^2  \\
&&+ \sum_{p \in \mathcal{P}} \sum_{m=1}^{M_p} \sum_{t=1}^{T} c_{p,m}^t(z_p^t-y^t_{p,m})^2 \\
&&+ \sum_{p \in \mathcal{P}} \gamma_p \sum_{t=2}^T (z_p^t-z_p^{t-1})^2 \\
&\text{s.t.}&  \quad  \sum_{s \in \mathcal{S}} \exp({-c_s}) = 1 
\end{eqnarray*}Note that in the warm-up period the soft sensors are constructed using the same methodology but the neighbor set is derived from the set of observed measurements in the time interval $[1:T]$. 

Since there are two sets of variables in the joint optimization problem, we apply the coordinate descent algorithm \cite{wright2015coordinate} to solve the problem. Specifically, we initialize $z_p^t=\frac{1}{|\mathcal{S}_p|} \sum_{s \in \mathcal{S}_p} x_s^t$, $\forall p \in \mathcal{P}, t \in [1:T]$. Then, we iteratively update sensor reliability scores and estimated ground truth of process states in two steps until the Euclidean distance between the estimated ground truth of process states between two consecutive iterations is less than a threshold:
\begin{eqnarray*}
\label{eq:converge}
\frac{1}{T} \sum_{t=1}^T ||\mathbf{z}^t_N - \mathbf{z}^t_{N-1}|| < \epsilon
\end{eqnarray*}where $\mathbf{z}^t_N$ denotes the estimated ground truth of process states at time $t$ in the $N$th iteration. Concretely, in the first step we fix the estimated ground truth of process states and update sensor reliability scores by the same method in Section~\ref{sec:score_update} as follows:
\begin{eqnarray*}
\label{eq:reliability2}
c_s =- \ln (f) \quad \forall s \in \mathcal{S}
\end{eqnarray*}
where
\begin{footnotesize}
\begin{equation*}
 f=\frac{\sum_{t=1}^T \Big\{ (z_p^t-x_s^t)^2 +\sum_{p' \in \mathcal{P} \setminus p} \sum_{m=1}^{M_{p'}} g^t_{p',m,s} (z_{p'}^t-y^t_{p',m})^2 \Big\} }{ \sum_{p \in \mathcal{P}} \sum_{\hat{s} \in \mathcal{S}_p} \sum_{t=1}^T \Big\{ (z_p^t-x_{\hat{s}}^t)^2 +  \sum_{p' \in \mathcal{P} \setminus p} \sum_{m=1}^{M_{p'}}  g_{p',m,\hat{s}}^t (z_{p'}^t-y^t_{p',m})^2 \Big\} } \nonumber
\end{equation*}
\end{footnotesize}
In the second step, we fix sensor reliability scores and estimate the ground truth of process states. Taking partial derivative of $L$ with respect to $z_p^t$ be zero, the ground truth of process states can be estimated by solving the following system of linear equations:
\begin{eqnarray*}
\label{eq:states1}
&& \sum_{s \in \mathcal{S}_p}c_s z_p^t + \sum_{m=1}^{M_p} c_{p,m}^t z_p^t + \nonumber\\
&&\mathbf{1}(t>1)\gamma_p (z_p^t- z_p^{t-1}) + \mathbf{1}(t<T)\gamma_p (z_p^t-z_p^{t+1})  \nonumber\\
&&= \sum_{s \in \mathcal{S}_p}c_s x^t_s + \sum_{m=1}^{M_p} c_{p,m}^t y_{p,m}^t   \quad \forall p \in \mathcal{P}, t \in [1:T] 
\end{eqnarray*}

\section{Implementation Issues}
\label{sec:implement}
In this section, we discuss some implementation issues of RelSen. First of all, since the value range for the states of distinct physical processes can be rather different, our framework will tend to assign higher reliability scores to sensors which monitor physical processes with smaller value range if the raw measurements are used. Therefore, we suggest to normalize the values of measurements into range [0,1] for all sensors. Secondly, there are a few hyperparameters to set up before running the algorithms in our framework. We illustrate the trade-off by setting different values of these hyperparameters:
\begin{itemize} 
\item $r$: With a smaller value of $r$, the derived soft sensors for a physical process will be less correlated, thus the possibility of having duplicated information sources for a physical process will be reduced. However, it will also increase the possibility of under-fitting for soft sensors, thus the soft sensors will provide less information in our optimization-based framework. A good way to tune $r$ is to apply cross validation on the data in the warm-up period.
\item $K$: With a large value of $K$, the neighborhoods may include too many training points that can result in regressions that oversmooth. Conversely, neighborhoods with too few points can result in regressions with incorrectly steep extrapolations \cite{gupta2008adaptive}. $K$ can also be tuned by cross validation on the data in the warm-up period. Furthermore, since deriving the K nearest neigbours from a time-series dataset with growing size can be time-consuming, we propose to derive the neighborhood from a fixed-size dataset where the points in the dataset are randomly sampled from the whole time series.
\item $\{M_p\}_{p \in \mathcal{P}}$: We construct different number of soft sensors for different physical processes. A good principle to decide $M_p$ is that a monitored process with less redundant sensors shall generally have more soft sensors.
\item $\{\gamma_p\}_{p \in \mathcal{P}}$: As mentioned, the value of $\gamma_p$ shall be decided by the prior knowledge about the smoothness of the monitored physical process.
\item $\epsilon$: The value of $\epsilon$ shall be close to zero. Assume the values of sensor measurements are normalized, we explicitly set $\epsilon=1 \times 10^{-5}$. We find that setting a smaller value will have limited impacts on the results.  
\item $l$: With a larger value of $l$, more data points will be considered for updating sensor reliability scores, thus the computed reliability scores will be smoother and we can have higher confidence in identifying unreliable sensors. However, in the meantime the computing cost is also increased by considering more data points and the latency of identifying unreliable sensors may also be increased. 
\item $T$: It is required that $T>\max(K,l)$. Moreover, $T$ should also be considerably larger than $K$ to ensure that the constructed soft sensors in the warm up period are fitted by different sets of data.
\end{itemize}

\section{Experiment on Sensors for Outdoor Air Pollution Monitoring}
\label{sec:airquality}
In this experiment, we apply RelSen for sensor reliability monitoring and data cleaning in an outdoor air pollution monitoring system. Specifically, in our experiment 16 sensors are deployed to monitor 6 physical processes in a small area. The 6 monitored physical processes are the concentrations of NO$_2$, NO, PM$_{10}$, PM$_{2.5}$, CO and O$_3$ in the air. Among the 16 deployed sensors, the number of sensors for monitoring each physical process is illustrated in Table~\ref{tab:schema_air}. Each sensor reports its measurement every hour. We collected measurements from the 16 sensors for four months.
 \begin{table}
\begin{center}
\setlength{\tabcolsep}{0.5em}
\def\arraystretch{1.1}
\begin{tabular}{ |c |c |c|c|c|c|c| }
\hline
 Monitored process & NO$_2$ & NO  & PM$_{10}$ & PM$_{2.5}$ & CO  & O$_3$ \\
\hline
Num. of sensors & 5 & 3 & 3 & 2 & 2 & 1\\
  \hline  
\end{tabular}
\end{center}
\caption{The sensor deployment schema in the outdoor air pollution monitoring experiment \label{tab:schema_air} }
\end{table}

 \begin{figure*}
\subfigure{
\includegraphics[width=.3\textwidth]{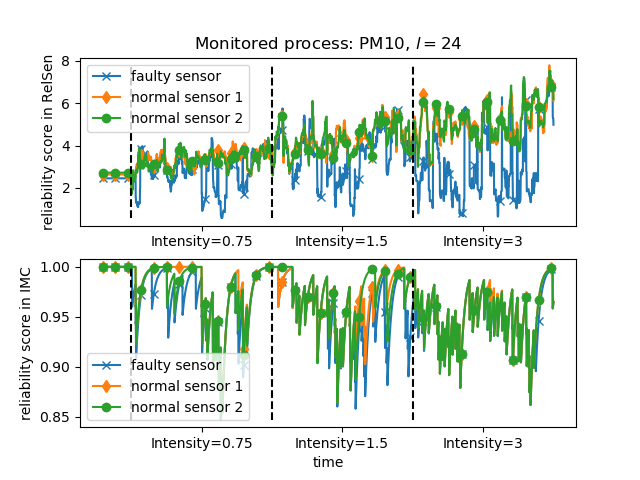}
}
\subfigure{
\includegraphics[width=.3\textwidth]{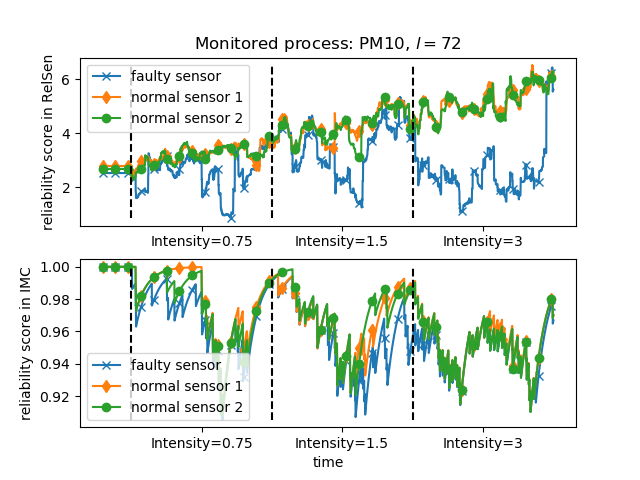}
}
\subfigure{
\includegraphics[width=.3\textwidth]{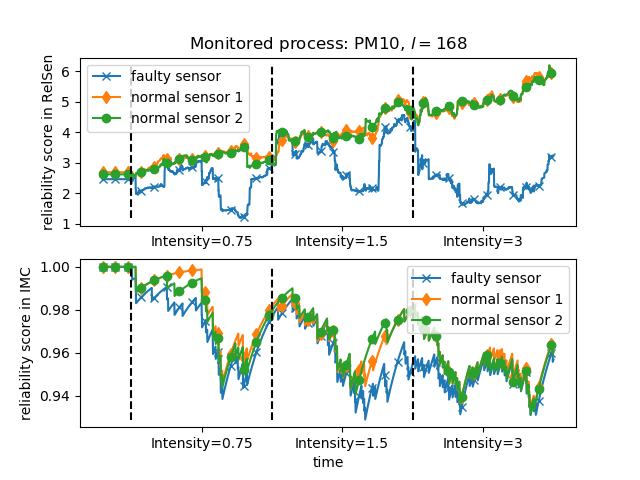}
}
\caption{Comparison of reliability score traces for sensors monitoring PM$_{10}$ under SHORT faults}
\label{fig:rel_short}
\end{figure*}

\begin{figure*}
\centering
\subfigure{
\includegraphics[width=.3\textwidth]{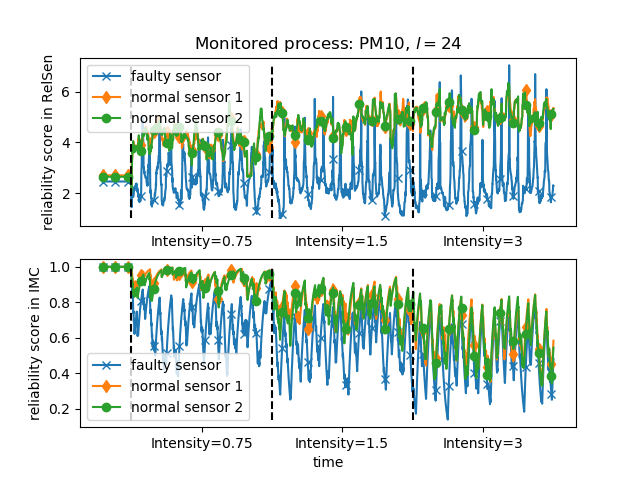}
}
\subfigure{
\includegraphics[width=.3\textwidth]{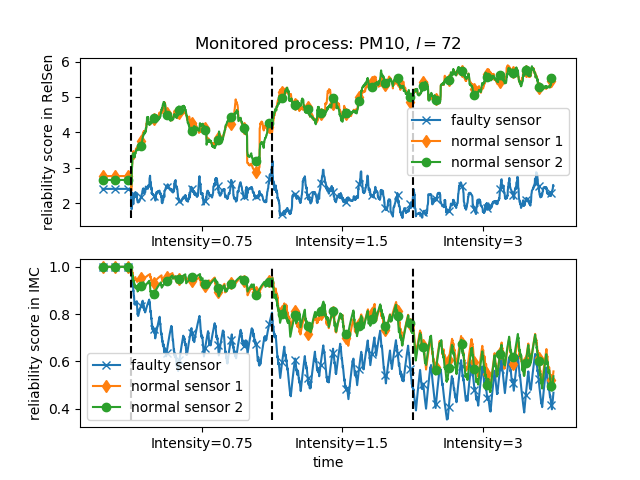}
}
\subfigure{
\includegraphics[width=.3\textwidth]{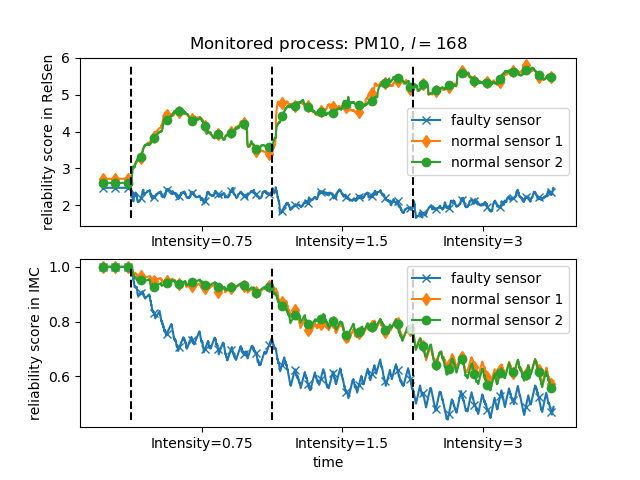}
}
\caption{Comparison of reliability score traces for sensors monitoring PM$_{10}$ under NOISE faults}
\label{fig:rel_noise}
\end{figure*}

\begin{figure*}
\centering
\subfigure{
\includegraphics[width=.3\textwidth]{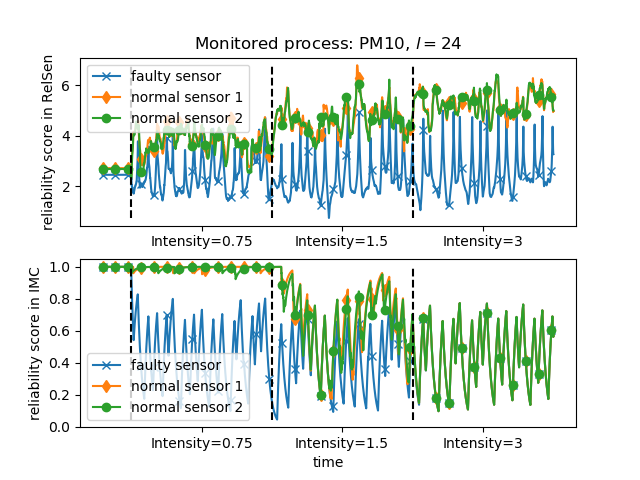}
}
\subfigure{
\includegraphics[width=.3\textwidth]{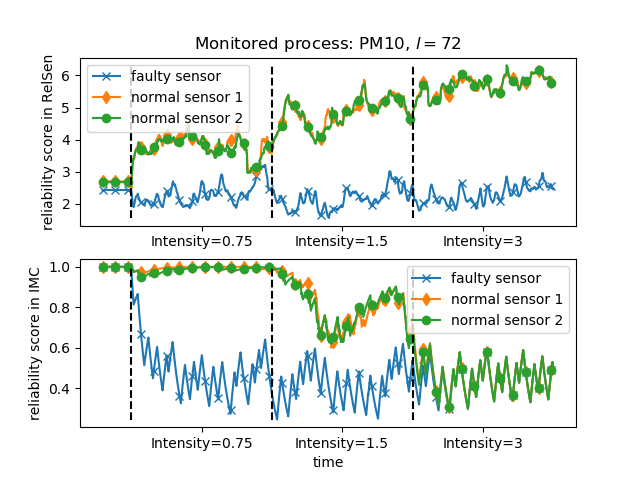}
}
\subfigure{
\includegraphics[width=.3\textwidth]{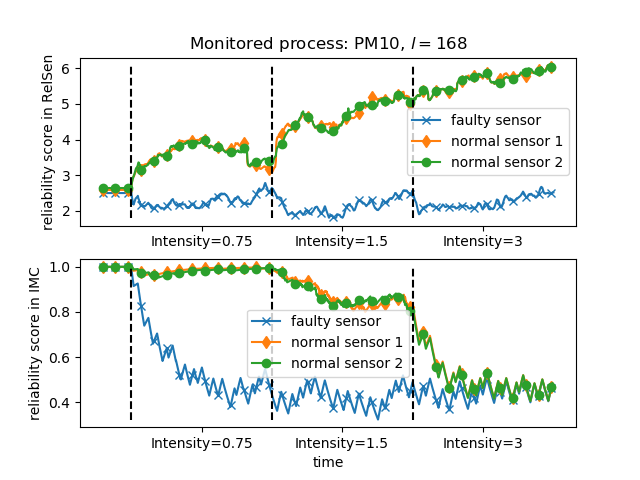}
}
\caption{Comparison of reliability score traces for sensors monitoring PM$_{10}$ under CONSTANT faults}
\label{fig:rel_constant}
\end{figure*}

Since the ground truth of process states is unknown, it is difficult for us to evaluate the performance of our method. Thus, we consider injecting artificial sensor data faults to the collected data. Specifically, we consider three types of sensor data faults which have been most commonly observed in real deployments as described in \cite{sharma2010sensor}: SHORT faults, NOISE faults and CONSTANT faults. For SHORT faults, there is a sharp change in the measurements between two successive points from a single sensor; For NOISE faults, the noise variance of the sensor increases within a number of successive data points; For CONSTANT faults, the sensor reports values with a constant offset for a number of successive data points. For evaluation purpose, we treat the mean value of measurements from the sensors co-monitoring a physical process before fault injection as the ground truth. In the implementation of our method, we set length of the warm-up period to seven days, thus $T=168$. We further set $r=0.7$, $K=48$, and $\{M_p\}_{p \in \mathcal{P}}$ are set to values such that each process has five hard and soft sensors in total, $\{\gamma\}_{p \in \mathcal{P}}$ are set to 1 for all processes. The length of sliding window $l$ is set to $24,72,168$ for experimental purpose.

\subsection{Baseline Methods}
To demonstrate the benefits of our method, we further compare our performance with four baseline data cleaning methods:
\subsubsection{MEDIAN}The MEDIAN method is commonly used in practice. It estimates a monitored process state as the median value of measurements from its responsible sensors. In case of only having two sensors, e.g., PM$_{2.5}$ and CO in this experiment, the MEDIAN method will take the mean as the median.
\subsubsection{MEAN}The MEAN method is also very commonly used in practice. It estimates a monitored process state as the mean value of measurements from its responsible sensors.
\subsubsection{IMC}Like RelSen, the IMC method is also a sensor reliability-based data cleaning method. It estimates a monitored process state as the weighted sum of the measurements from its responsible sensors, such that $z^t_p = \frac{\sum_{s \in \mathcal{S}_p} c_s^t x^t_s }{\sum_{s \in \mathcal{S}_p} c_s^t}$, where the weights are decided by the sensor reliability scores updated by the following rule:
\[ c_s^t = \frac{1}{l}\sum_{k=t-l-1}^{t-1} cons(x^k_s)
\]
and 
\[ cons(x^t_s) =
  \begin{cases}
    1       & \quad \text{if } |x_s^t-z^t_p| \leq tol  \\
    0 & \quad \text{otherwise}
  \end{cases}
\]where $l$ is the length of a sliding window, $tol$ is an error threshold below which a sensor measurement is regarded as consistent with the estimated process state. We refer more details of the IMC method to \cite{zhang2014cleaning}. Note that in case of no redundancy such as O$_3$ in the experiment, the IMC method will deteriorate to reporting the measurements from the sensor without data cleaning. In our experiment, $tol$ is tuned to $0.05$, $l$ is also set to $24,72,168$ for comparison.
\subsubsection{BayesGMM}BayesGMM is a parameter estimation method which employs a Bayesian framework such that $ \mathbf{x} = \mathbf{z} + \bm{\varepsilon} $, where $\mathbf{x}$ and $\mathbf{z}$ are vectors of random variables representing sensor measurements and process states respectively, $\bm{\varepsilon}$ is a Gaussian random vector with zero mean and diagonal covariance matrix $\bm{\Theta}$. In this method, a training stage is required in which no sensor faults occur, and the distribution of $\mathbf{z}$ is initialized via the Gaussian mixture model (GMM):
\[ P(\mathbf{z} ) = \sum_{s=1}^K P(\mathbf{z} | s) P(s) \]where $s$ is the label for the $s$th mixture component. During the monitoring stage, the process states are estimated as \[ \mathbf{z}^t = E(\mathbf{z} | \mathbf{x}^t, \bm{\Theta}^t ) \] where an Expectation-Maximization (EM) algorithm \cite{dempster1977maximum} is used to estimate $\mathbf{z}^t$ and $\bm{\Theta}^t $ simultaneously. More details of the BayesGMM method can be found in \cite{yuan2007bayesian}. Note that the BayesGMM method does not explicitly utilize sensor redundancy for process estimation, as a result, let $z_s^t$ denote the estimated process state for sensor $s$, we calculate $z_p^t = \sum_{s \in \mathcal{S}_p} z_s^t$. Furthermore, in the experiments, the length of the training stage is set to $T$, and the number of GMM components is chosen by minimizing the Bayesian information criterion (BIC) score \cite{keribin2000consistent}.

It is worth noting that in our experiments, we assume the state transition dynamics of the physical processes cannot be predefined, thus methods which require prior knowledge on state transition dynamics of the physical processes such as dynamic state-space models \cite{petris2009dynamic,zhang2017state} are not considered in our context.

\subsection{Fault Injection}
To check the performance of each method under different fault types, the experiment under each type of faults is conducted separately. Specifically, to inject a particular fault type, one responsible sensor for each physical process is selected as the faulty sensor. Furthermore, we evenly divide the data after the warm-up period into three stages, namely low, medium and high intensity stages. We inject faults with intensities $f=\{0.75,1.5,3\}$ in the three stages respectively. 

To inject SHORT faults, we randomly pick $5\%$ of data points for each faulty sensor and replace $x_s^t$ with $\hat{x}_s^t$ such that $\hat{x}_s^t = x_s^t + f \times x_s^t$. To inject NOISE faults for a faulty sensor, we replace $x_s^t$ with $\hat{x}_s^t = x_s^t + \mathcal{N}(0,f\times\sigma_s^2)$ with a random duration from 10 to 50 data points such that adjacent contaminated segments are 24 data points away from each other, where $\sigma_s$ is the standard deviation of sensor $s$ in the data. CONSTANT faults are injected in a similar way with the NOISE faults, the only difference is that we replace $x_s^t$ with $\hat{x}_s^t = x_s^t + f\times\sigma_s$.

\begin{table*}
\caption{The mean absolute errors of cleaned sensor data compared with ground truth in the air pollution monitoring system.\label{tab:airmae}}
\begin{center}
\setlength{\tabcolsep}{0.5em}
\def\arraystretch{1.1}
\begin{tabular}{ |c|c |c |c|c|c|c|c|c|c|l|}
\cline{1-10} 
 & \multirow{2}{*}{MEDIAN} & \multirow{2}{*}{MEAN} & \multirow{2}{*}{BayesGMM} & \multicolumn{3}{c|}{IMC} & \multicolumn{3}{c|}{RelSen}   \\
\cline{5-10}
 &  &  &  & $l=24$ &$l=72$&$l=168$ & $l=24$ &$l=72$&$l=168$   \\
\hline 
\multirow{3}{*}{NO$_2$} & \textbf{0.004} & 0.081 & 0.044 & 0.08 & 0.08 & 0.08 & 0.049 & 0.045 & 0.043 & under SHORT faults  \\
 & \textbf{0.004} & 0.049 & 0.026 & 0.031 & 0.033 & 0.033 & 0.022 & 0.018 & 0.018 &  under NOISE faults   \\
 & \textbf{0.004} & 0.063 & 0.026 & 0.047 & 0.048 & 0.046 & 0.029 & 0.024 & 0.024 &  under CONSTANT faults  \\
\hline
\multirow{3}{*}{NO} & \textbf{0.003} & 0.016 & 0.015 & 0.016 & 0.016 & 0.016 & 0.014 & 0.014 & 0.014 &  under SHORT faults   \\
 & \textbf{0.003} & 0.039 & 0.015 & 0.033 & 0.034 & 0.034 & 0.02 & 0.019 & 0.019 &  under NOISE faults  \\
 & \textbf{0.004} & 0.047 & 0.034 & 0.039 & 0.039 & 0.038 & 0.03 & 0.028 & 0.029 & under CONSTANT faults  \\
\hline 
\multirow{3}{*}{PM$_{10}$} & \textbf{0.01} & 0.139 & 0.073 & 0.139 & 0.138 & 0.138 & 0.059 & 0.051 & 0.05 &  under SHORT faults  \\
 & \textbf{0.009} & 0.069 & 0.036 & 0.055 & 0.058 & 0.058 & 0.03 & 0.027 & 0.027 &  under NOISE faults   \\
 & \textbf{0.01} & 0.087 & 0.043 & 0.073 & 0.071 & 0.067 & 0.039 & 0.033 & 0.033 &  under CONSTANT faults  \\
\hline 
\multirow{3}{*}{PM$_{2.5}$} & 0.156 & 0.156 & 0.052 & 0.156 & 0.156 & 0.156 & 0.056 & 0.054 & \textbf{0.051} &  under SHORT faults  \\
 &0.098 & 0.098 & 0.04 & 0.098 & 0.098 & 0.098 & 0.036 & \textbf{0.033} & \textbf{0.033} &   under NOISE faults  \\
 & 0.126 & 0.126 & \textbf{0.05} & 0.126 & 0.126 & 0.126 & 0.058 & 0.053 & 0.052 & under CONSTANT faults  \\
\hline 
\multirow{3}{*}{CO} & 0.271 & 0.271 & \textbf{0.047} & 0.271 & 0.271 & 0.271 & 0.069 & 0.066 & 0.063 &  under SHORT faults  \\
 &0.087 & 0.087 & 0.046 & 0.087 & 0.087 & 0.087 & 0.031 & 0.029 & \textbf{0.028} &   under NOISE faults   \\
 &0.107 & 0.107 & 0.059 & 0.107 & 0.107 & 0.107 & 0.056 & \textbf{0.052} & \textbf{0.052} &  under CONSTANT faults \\
\hline 
\multirow{3}{*}{O$_3$} & 0.644 & 0.644 & 0.135 & 0.644 & 0.644 & 0.644 & \textbf{0.109} & 0.11 & 0.11 &  under SHORT faults  \\
 &0.315 & 0.315 & 0.146 & 0.315 & 0.315 & 0.315 & \textbf{0.128} & 0.131 & 0.13 &    under NOISE faults   \\
 & 0.402 & 0.402 & 0.145 & 0.402 & 0.402 & 0.402 & \textbf{0.108} & 0.116 & 0.117 &  under CONSTANT faults  \\
\hline
\multirow{3}{*}{Avg.} & 0.181&0.218&0.061&0.218&0.218&0.218&0.059&0.057&\textbf{0.055}& under SHORT faults  \\
 &0.086&0.11&0.051&0.103&0.104&0.104&0.045&0.043&\textbf{0.042}&   under NOISE faults   \\
 & 0.109&0.139&0.06&0.132&0.132&0.131&0.053&0.051&\textbf{0.051}&  under CONSTANT faults  \\
\hline 
\end{tabular}
\end{center}
\end{table*}

\subsection{Performance Evaluation}
To illustrate the effectiveness of RelSen on timely identifying unreliable sensors, we compare the traces of reliability scores for sensors generated by RelSen and IMC under SHORT, NOISE and CONSTANT faults in Figure~\ref{fig:rel_short}, \ref{fig:rel_noise} and \ref{fig:rel_constant} respectively. Due to lack of space, we select the PM$_{10}$ sensors as representative cases for illustration. As shown in the figures, we can easily distinguish the faulty sensors from normal sensors based on the reliability scores generated by RelSen under all fault types. Specifically, we can observe a downward trend of reliability scores for the faulty sensor and an upward trend for the normal sensors (the trends become more clear with larger $l$). Since the reliability scores in RelSen are evaluated relatively, the gap between the reliability scores from normal and faulty sensors can be rather obvious. Consequently faulty sensors are easy to identify by monitoring the reliability scores. On the contrary, we find that the IMC method cannot identify the faulty sensor under SHORT faults because the method uses the proportion of identified abnormal measurements to set the reliability score, and thus its ability to detect sensors with SHORT faults (only appear one data point each time) is very limited. Moreover, when the intensity of injected NOISE and CONSTANT faults is high, all sensors will be treated as unreliable sensors using the reliability scores generated by IMC. This is because when the measurement error of a sensor becomes too large, the distance of all sensors' measurements to the weighted mean will all go beyond the threshold value. Notably, although sensor measurement errors are assumed to be generated by sensor faults in our experiments, they can also be generated by intrinsic sensor noise caused by manufacturing imperfection. RelSen can also identify the different level of intrinsic noise for sensors. The evidence is that the sensors are assigned with different reliability scores by RelSen in the warm-up period during which no sensor faults has been injected.

To evaluate the performance of RelSen on data cleaning, we summarize the mean absolute errors (MAEs) of the cleaned sensor data generated by different methods under the three fault types in Table~\ref{tab:airmae} (the MAEs are calculated using the normalized values). We can see that RelSen achieves the best average accuracy under all the three fault types. Furthermore, although RelSen can achieve different accuracy with different values of $l$, the difference is generally minor compared with the gap with other methods.
Additionally, the performance of RelSen is also more robust compared with the MEDIAN, MEAN and IMC methods which exhibit very different accuracy in different scenarios, i.e., different processes under monitoring, different number of redundant sensors, different sensor fault types. BayesGMM has comparable overall accuracy with RelSen in this experiment, however, there is a drawback that the parameter estimation method requires the data in the training stage to capture the profile of the system dynamics. In the next experiment, we show that the performance of parameter estimation methods will deteriorate dramatically when the requirement cannot be satisfied.

\section{Experiment on Sensors for Condition Monitoring in a Cement Rotary Kiln}
\label{sec:kiln}
In this section, we present our results of experiment conducted on a sensor-based condition monitoring system in a cement rotary kiln. Specifically, in the system there are 20 sensors in total among which 16 monitor the temperature and negative pressure on the inlet and outlet cones of four cyclones deployed in parallel, one monitors the negative pressure on the outlet of a connected decomposition furnace, three monitor the temperature on the bottom, middle and top of the decomposition furnace. Each sensor reports its measurement every 30 seconds. We collected measurements from the 20 sensors for a week.

\begin{figure}[b]
  \centering
  \includegraphics[width=.75\linewidth]{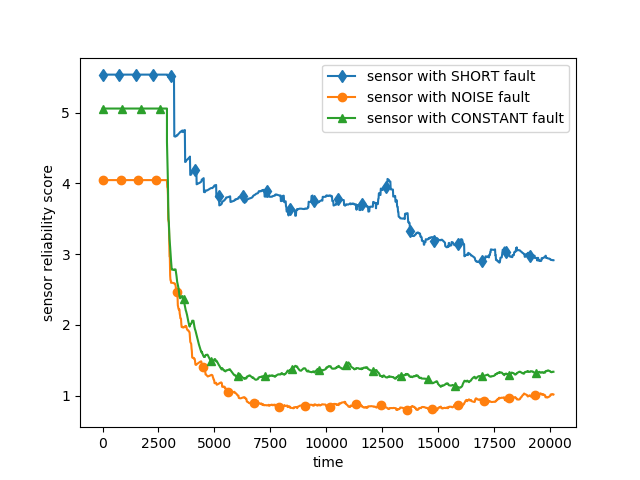}
\caption{The reliability scores of the faulty sensors generated by RelSen in the condition monitoring system of the cement rotary kiln}
  \label{fig:kiln_rel}
\end{figure}

\begin{figure}
\centering
\subfigure{
\includegraphics[width=.85\linewidth]{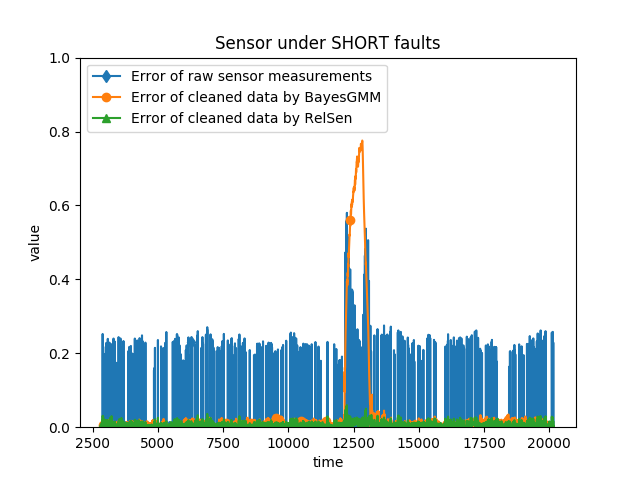}
}
\subfigure{
\includegraphics[width=.85\linewidth]{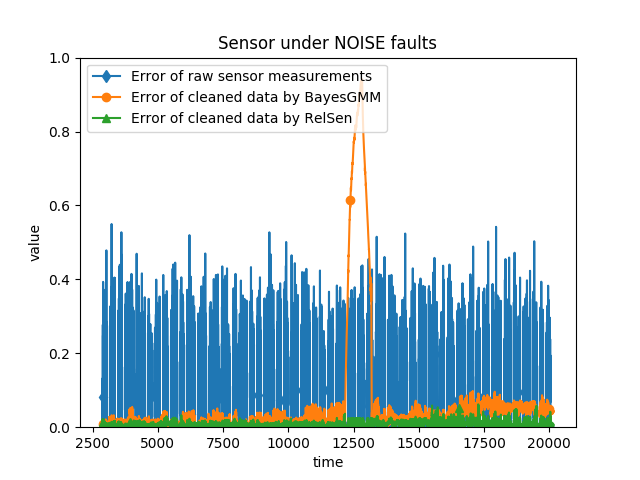}
}
\subfigure{
\includegraphics[width=.85\linewidth]{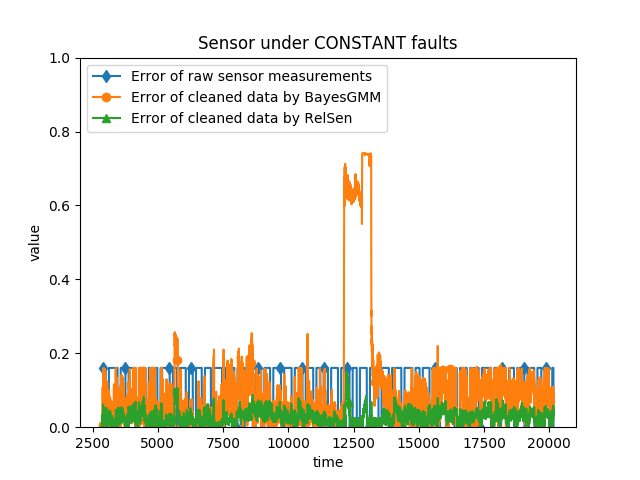}
}
\caption{Absolute error of sensor measurements and cleaned data from faulty sensors in the condition monitoring system of the cement rotary kiln}
\label{fig:kiln_ground}
\end{figure}

%\begin{table}
%\caption{The mean absolute errors between cleaned sensor data and ground truth in the condition monitoring system of the cement rotary kiln\label{tab:maekiln}}
%\begin{center}
%\begin{tabular}{ |c|c |c |}
%\hline 
% & BayesGMM & RelSen\\
%\hline
%Under SHORT fault  & 0.01 & 0.1 \\
%\hline
%Under NOISE fault & 0.01 & 0.1\\
%\hline
%Under CONSTANT fault  & 0.01 & 0.1\\
%\hline  
%\end{tabular}
%\end{center}
%\end{table}

Since each sensor monitors a distant physical process (temperature and negative pressure on different parts) in this experiment, the effects of MEDIAN, MEAN and IMC methods deteriorate to the same as the raw sensor measurements. Therefore, we mainly compare our data cleaning accuracy with BayesGMM. In the implementation of BayesGMM, we use the first day's data as the training data. 
We also set the length of the warm-up period in RelSen to one day, thus $T=2880$. We further set $r=0.7$, $K=48$, and $\{M_p\}_{p \in \mathcal{P}}=5$. $\{\gamma\}_{p \in \mathcal{P}}$ are set to 1 for all processes. The length of sliding window $l$ is set to $288$.

Three sensors, namely the temperature sensor on the inlet cone of Cyclone 4, the temperature sensor on the outlet cone of Cyclone 1, and the negative pressure sensor on the outlet of the decomposition furnace are selected as the faulty sensors with SHORT, NOISE and CONSTANT fault injection respectively. All the three fault types are injected together after the first day with intensity $f=1$ following the same equations in the previous experiment. For SHORT faults injection, $2\%$ of data points are contaminated. NOISE and CONSTANT faults are injected with a random duration of 240 to 360 data points such that adjacent contaminated segments are 120 data points away from each other.

In Figure~\ref{fig:kiln_rel}, we show the reliability scores of the faulty sensors generated by RelSen in the experiment. We observe that the reliability scores of the three faulty sensors decrease steeply after the warm-up period. The reliability scores are stabilized to values which are significantly lower than their values in the warm-up period after a certain length of time period. This shows that we can quickly identify the three faulty sensors by monitoring their reliability scores.

In Figure~\ref{fig:kiln_ground}, we present the absolute errors of raw sensor measurements and cleaned data from RelSen and BayesGMM for the three faulty sensors. The result shows that RelSen can remove a large proportion of sensor measurement errors and can achieve higher accuracy than BayesGMM under all the three fault types. We believe the main reason behind this is that the performance of BayesGMM deteriorates dramatically when the process characteristics change during the monitoring stage. Consequently, the trained model cannot capture the behavior of the physical process any more. The evidence is clear in Figure~\ref{fig:kiln_ground}, where BayesGMM fails to capture the dynamics of the physical processes around time points between 12000 to 13000. However, in RelSen we use random local linear regression for soft sensor construction, this allows the soft sensors to be promptly adapted to capture changing process characteristics, thus the estimated ground truth of process states are more accurate.

\section{Conclusion}
\label{sec:conclude}
In this paper, we have proposed RelSen: a novel optimization-based framework for simultaneous sensor reliability monitoring and data cleaning in sensor systems. The main logic behind RelSen is fairly straightforward: more reliable sensors should provide more accurate measurements; the ground truth of monitored process states should be closer to the measurements from more reliable sensors. By utilizing the cross correlation between multiple processes, RelSen can dynamically update the reliability scores of sensors and accurately clean sensor data in real time only given the measurements from sensors. In our experiments conducted respectively on sensor systems for outdoor air quality monitoring and cement rotary kiln condition monitoring, we demonstrated that RelSen can accurately and promptly identify unreliable sensors under three types of commonly observed sensor faults. Furthermore, we showed that RelSen outperformed several baseline methods regarding to data cleaning.

With less assumptions and knowledge requirements about the monitored process dynamics, we see potential for application of RelSen to a wide range of use-cases in the sensor-based IoT applications. In the future, we will study the potential to extend RelSen to a broader class of problems such as time-series data cleaning and model fusion in time-series ensemble learning.

\bibliographystyle{ACM-Reference-Format}
\bibliography{trustbib}

\end{document}